\documentclass[showpacs,amsmath,amssymb,prl,superscriptaddress,floatfix,twocolumn]{revtex4}
\usepackage{graphicx}
\usepackage{dcolumn}
\usepackage{bm}
\usepackage{color}

\begin{document}

\title{Evolutionary phase space in  driven elliptical billiards} 
\author{F. Lenz}
\email[]{lenz@physi.uni-heidelberg.de}
\affiliation{Physikalisches Institut,  Universit\"at Heidelberg,
Philosophenweg 12, 69120 Heidelberg, Germany}
\author{C. Petri}
\email[]{petri@physi.uni-heidelberg.de}
\affiliation{Physikalisches Institut,  Universit\"at Heidelberg,
Philosophenweg 12, 69120 Heidelberg, Germany}
\author{F. R. N. Koch}
\email[]{florian@physi.uni-heidelberg.de}
\affiliation{Physikalisches Institut,  Universit\"at Heidelberg,
Philosophenweg 12, 69120 Heidelberg, Germany}
\author{F. K. Diakonos}
\email[]{fdiakono@phys.uoa.gr}
\affiliation{Department of Physics, University of Athens, GR-15771 Athens,
Greece}
\author{P. Schmelcher}
\email[]{peter.schmelcher@pci.uni-heidelberg.de}
\affiliation{Physikalisches Institut,  Universit\"at Heidelberg,
Philosophenweg 12, 69120 Heidelberg, Germany}
\affiliation{Theoretische Chemie, Physikalisch-Chemisches Institut,
Universit\"at Heidelberg, Im Neuenheimer Feld 229,
 69120 Heidelberg, Germany}
\date{\today}

\begin{abstract}
We perform the first long-time exploration of the classical dynamics of a
driven billiard with a four dimensional phase space. With increasing velocity of the ensemble we observe an evolution from a large chaotic sea
with stickiness due to regular islands to thin chaotic channels with diffusive motion leading to Fermi acceleration. As a surprising consequence, we encounter a crossover, which is not parameter induced but rather occurs dynamically, from amplitude dependent tunable subdiffusion to universal normal diffusion in momentum space. In the high
velocity case we observe particle focusing in phase space.
\end{abstract}
\maketitle

Billiards are a widespread paradigm  to study complex dynamics in many areas of
physics. Two dimensional systems are especially appealing, since unlike one
dimensional ones they possess already a rich phase space structure, which
can range from integrable over mixed to fully chaotic \cite{Berry:1982}. Billiard systems can
be realized experimentally in different ways, such as using semi-conductor
heterostuctures \cite{Weiss:1991}, quantum dots \cite{Marcus:1992}, atom optical
\cite{Milner:2001} and  (superconducting) microwave setups
\cite{Stockmann:1999,Richter:2001}. Very recent applications of their 
dynamics are the investigation of directed emission from laser-microcavities
\cite{Gmachl:1998, Creagh:2007, Lee:2008} and the design of improved thermoelectric
efficiency, where billiards are used to tailor the desired microscopic properties 
\cite{Casati:2008}.

A natural generalization are time-dependent billiards, where the
boundary is driven, for example harmonically. This allows the study of
non-equilibrium processes such as Fermi acceleration (FA) \cite{Fermi:1949,
Lieberman:1972,
Karlis:2006}. There are only few studies of such two dimensional (2D) driven
billiards
\cite{Koiller:1995, Loskutov:1999, Kamphorst:1999, Kamphorst:2007,
Gelfreich:2008, Lenz:2007, Lenz:2008}, since the long-time numerical iteration of the
corresponding four dimensional (4D) implicit mapping is computationally challenging
and in particular the resulting phase space is hard to analyze  due to its high dimensionality and
unboundedness.  So far this prevented any investigation
addressing the long-time behavior of corresponding dynamical quantities. Key questions are whether there will be FA \cite{Fermi:1949}, i.e., unbounded
energy gain of an ensemble exposed to time-dependent external potentials and more specifically  what laws this FA will follow depending on the geometry and the driving of the billiard.  In one
dimension, the existence of FA depends exclusively  on the driving law \cite{Lichtenberg:1992}, in 2D
the situation is much more complicated, since the dynamics of the corresponding
static systems can be integrable, mixed or chaotic. Recently  the
existence of FA in the harmonically driven elliptical billiard was demonstrated \cite{Lenz:2008}, augmenting the ``LRA-conjecture''  which says that only time-dependent systems with a mixed or chaotic static
counterpart exhibit FA \cite{Loskutov:1999}. However, none
of the mentioned works analyzed the full 4D phase space and
investigated how its structure determines the diffusion in momentum space,
especially  with respect to the long time behavior of the diffusion and acceleration process. 

In this Letter, we present, for the first time, a detailed exploration
of the long-term dynamics of a driven billiard with a 4D phase space. It
is shown that the phase space geometry evolves, with increasing velocity
of the propagating ensemble, from a large chaotic sea with rich stickiness
structures due to regular islands, to large dominating regular domains
with thin chaotic acceleration channels. This change in the phase space
structure leads to a crossover from amplitude dependent tunable
subdiffusion to universal normal diffusion in momentum space. Due to its
acceleration with increasing number of collisions the particle ensemble
propagates effectively in an evolutionary phase space which we study in
depth.

The geometry of
the time-dependent elliptical billiard is  given by
\begin{equation}\label{eq:EllBoundary}
 \left( \begin{array}{c} x(t) \\ y(t) \\ \end{array}\right)=  \left
(\begin{array}{c} a(t)
\cos \phi \\
   b(t)\sin \phi\\ \end{array}\right )=\left
(\begin{array}{c} (A_0 + C \sin (\omega t))
\cos \phi \\
   (B_0 + C \sin (\omega t)) \sin \phi\\ \end{array}\right ),
\end{equation}
where $(x(t),y(t))$ is a point on the boundary, $\phi$ is the  corresponding elliptic polar angle, $C>0$ is the driving
amplitude, $\omega$ is the frequency of oscillation and $A_0$, $B_0$ are the
equilibrium values of the long and the short half-diameter, respectively. We fix
$\omega=1$, $A_0=2$ and $B_0=1$ without loss of generality (arbitrary units). Due to ballistic motion in between collisions, the dynamics can be
described by an implicit $4D$ map \cite{Lenz:2007} using the variables
$(\xi_n,\phi_n, \alpha_n, v_n)$, where $v_n=\vert \bm{v}_n \vert$
is the modulus of the particles velocity,  $\xi_n = \omega t \mod 2\pi$ is
the phase of the boundary oscillation and $\alpha_n$ is the angle between
$\bm{v}_n$ and the tangent on the boundary at the $n$-th collision. Note that
the corresponding static ellipse is integrable \cite{Berry:1982}, since besides the
energy the product of the angular momenta around the two foci
\begin{equation}\label{eq:F}
    F(\phi,\alpha)= \frac{\cos^2\alpha \cdot(1+ (1-\varepsilon^2) \cot^2
\phi)-\varepsilon^2}{1+ (1-\varepsilon^2) \cot^2 \phi -
    \varepsilon^2},
\end{equation}
where $\varepsilon$ is the eccentricity, is conserved. The static phase space is
globally divided by the
separatrix ($F=0$), connecting the two hyperbolic fixed points, into rotators
($F~>~0$) and librators ($F~<~0$). Two elliptic fixed points are
located at the  minima $F_{min}=-\varepsilon^2/(1-\varepsilon^2)$. Periodic orbits with short periods are located around the separatrix, whereas the existence of period orbits in the librator region close to the elliptic fixed points requires comparably long periods \cite{Koiller:1995}. This will later be important for the understanding of the distribution of periodic orbits in the driven case. 

In ref. \cite{Lenz:2008} we showed that the driven ellipse exhibits Fermi acceleration
(FA). Here we present a comprehensive analysis of the $4D$ phase space,
specifically we explore
how its composition changes with increasing $v$ and demonstrate how this affects the
long-time evolution of FA in the system. We first analyze the periodic orbits and their stability and consider subsequently the phase space density $\rho$ of a typical ensemble in detail, by exploring its dependence on the number of collisions $n$ as well as the velocity $v$. Due to the occurrence of FA, the phase
space is open with respect to $v$. Unlike in the static system, $\alpha$ is not restricted to
$[0, \, \pi]$, but can now range from $0$ to $2\pi$ \cite{Koiller:1995}.
Nevertheless, the total phase space is not just given by the $4D$ cuboid
$[0,\,2\pi] \times [0,\,2\pi] \times[0,\,2\pi]\times[0, \, \infty]$ spanned by
$\xi,\,\phi,\,\alpha$ and $v$, but possesses a more complicated topology, 
since e.g. for $\xi\in [\pi/2,\, 3\pi/2]$ (the ellipse is contracting) $\alpha$
is restricted to $\alpha \in [0,\, \pi]$ \cite{Petri:2008}. 
\begin{figure}[htp]
\centerline{\includegraphics[width=\columnwidth]{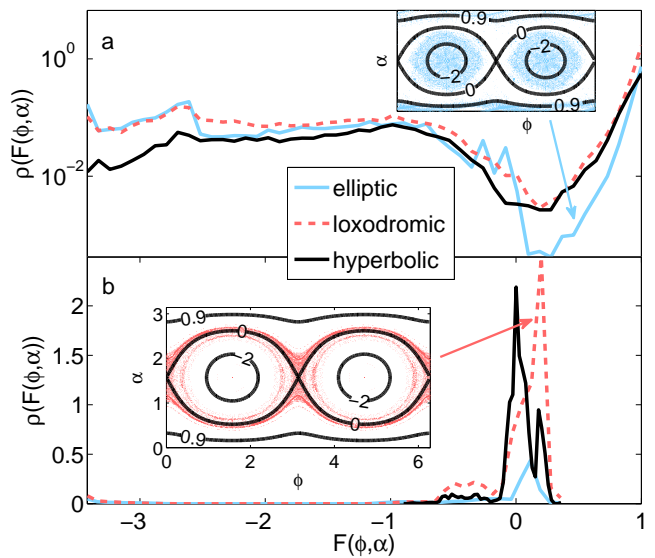}}
\caption{(Color online) Density of periodic orbits up to period 100 as a function of $F$ with $v_1<v<v_2$ and $v_1=0,\,v_2=1$ in a) and $v_1=30,\,v_2=40$ in b).  The $F(\phi,
\alpha)\approx0$ region is depleted for low velocities (a), and is populated
predominantly for higher velocities (b). Exemplary, the distribution of periodic orbits is shown in the insets in the $\phi\times\alpha$ plane, where the black curves are the $F=const.$ isolines.} \label{fig:fig1}
\end{figure}
\begin{figure}[hbp]
\centerline{\includegraphics[width=\columnwidth]{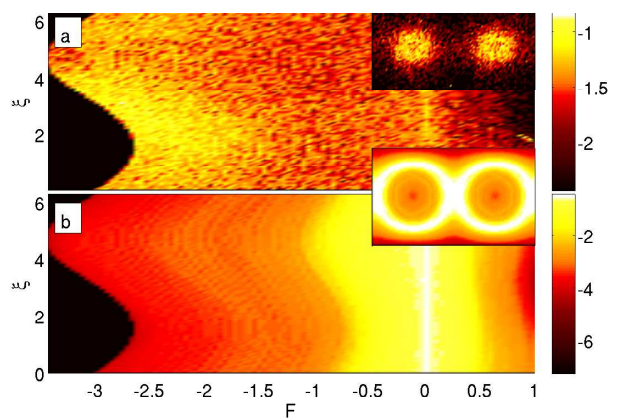}}
\caption{(Color online) Collision resolved phase space density $\rho_{n_1,n_2}(F,\xi)$ (log colormap), $n_1=10,\,n_2=10^2$ (a) and $n_1=4\cdot10^6,\,n_2=10^7$ (b). $\rho_{n_1,n_2}(F,\xi)$ is largest around $F\approx-2.5$ for small $n$ (a), whereas the $F(\phi, \alpha)\approx0$ region is predominantly populated  for high collision numbers (b). In the insets, the corresponding $\rho_{n_1,n_2}(\phi,\alpha)$ are shown.} \label{fig:fig2}
\end{figure}

Finding
 (unstable) periodic orbits in such a high dimensional phase space is an intricate task and becomes more difficult with increasing velocity of the orbits, since the typical period grows linearly with $v$. We use a variant of the method developed in \cite{Schmelcher:1997}. Since the mapping of the
driven ellipse is $4D$ and area-preserving, there are three possible
types of eigenvalues of the corresponding real monodromy matrix: (i) Four complex eigenvalues (elliptic), (ii) two real and two complex
eigenvalues (loxodromic) and (iii) four real eigenvalues (hyperbolic). The first case corresponds to
stable \footnote{In principle one has to distinguish in higher
dimensions between spectral and linear stability \cite{Lichtenberg:1992},
however this does not play a role here.}, whereas the latter two represent
unstable fixed points \cite{Lichtenberg:1992}. Performing extensive numerical computations we detected the periodic orbits up to period 100. \color{black} The density $\rho(F)$ of the periodic orbits as a function of $F(\phi,\alpha)$ is shown in Fig.~\ref{fig:fig1}. By using $F$, we can effectively map the $\phi\times\alpha$ plane onto a single dimension (see the insets of Fig.~\ref{fig:fig1} where the $F=const.$ isolines are shown in black). Fig.~1 shows periodic orbits with $v_1 < v < v_2$ for (a) $v_1=0$,
$v_2=1$ and (b) $v_1=30$, $v_2=40$. For small values of $v$ (see Fig. \ref{fig:fig1}a), the
periodic orbits up to period 100 cover the whole available $F$-range, and thus the whole $\phi\times\alpha$ plane, except for a narrow region $F\approx0$ around the
separatrix  of the static system. This has to be expected,
since the collisional dynamics with the oscillating walls deviates the  most for low $v$ from the dynamics of the corresponding static system. The region of the separatrix is associated with the highest instability and in particular the stable periodic orbits existing in the static case around $F\approx 0$ have been destroyed. The opposite scenario can be seen for high values of $v$, see
Fig.~\ref{fig:fig1}b. Here we expect the fingerprints of the emergence of a quasistatic regime for the billiard: The particles are fast enough to experience many collisions with the oscillating walls within a narrow interval of the phase of oscillation. The relative momentum change obeys $\delta v/v \ll 1$ at each collision and the angle of incidence is almost the angle of reflection. As a result they trace orbits of the static system. As can be seen in Fig.~\ref{fig:fig1}b, the region $F\approx0$ possesses the highest density of period orbits, in particular unstable ones (hyperbolic and loxodromic), whereas the librator region around the two elliptic fixed 
points of the static system is depleted. 

The consequences of the above described changes in the distributions of the
periodic orbits with increasing $v$ become immediately clear when studying the
evolution (in terms of the number of collisions) of the phase space density
$\rho_{n_1,n_2}(F,\xi)=\frac{1}{n_2-n_1+1}\sum_{n=n_1}^{n_2}\int_v \rho(F,\xi,v,n)dv$ of an ensemble 
\begin{equation}
 \rho(F,\xi,v,n) = \frac{1}{N_p}
\sum_{i=1}^{N_p}
\delta(\xi-\xi_i^n)\delta(F-F(\phi_i^n,\alpha_i^n))\delta(v-v_i^n), 
\end{equation}
where the indices stand for the $i$th particle at collision number $n$,
respectively. An  ensemble of $N_p=10^3$  particles, with
$v_0=0.1$ and $\xi_0,
\phi_0, \alpha_0$ distributed uniformly and randomly ($\xi_0$, $\phi_0\in[0,
\,2\pi]$, $\alpha_0\in[0, \, \pi]$), is iterated for $10^7$ collisions. The  visiting probability  is shown in a log scale colormap in Fig.~\ref{fig:fig2}. In the beginning ($n_1=10$,
$n_2=100$) of the evolution,  the region around the
elliptic fixed points of the static system possesses the highest visiting
probability, see Fig.~\ref{fig:fig2}a and especially the inset, where the corresponding $\rho_{n_1,n_2}(\phi,\alpha)$ is shown. For large collisions numbers ($n_1=4\cdot 10^6$, $n_2=10^7$), see
Fig.~\ref{fig:fig2}b and the inset, the ensemble is located predominantly in a region around
the separatrix ($F(\phi,\alpha)=0$) of the corresponding static system. In the insets, the half-space $\alpha>\pi$ is not shown, since $\rho(\phi,\alpha)\approx0$ there. 
\begin{figure}[htp]
\centerline{\includegraphics[width=\columnwidth]{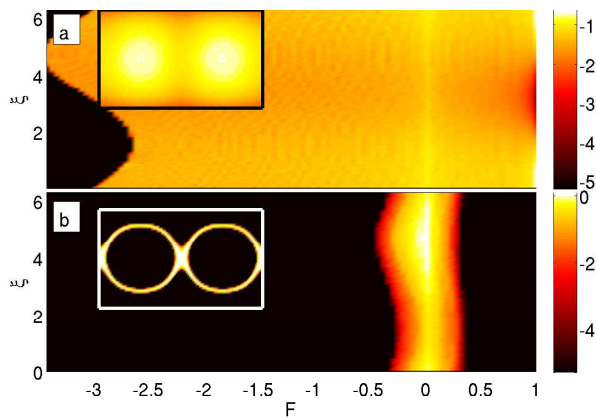}}
\caption{(Color online) Velocity resolved phase space density $\rho_{v_1,v_2}(F,\xi)$ (log colormap). For low velocities (a), the ensemble covers the whole
$F\times\xi$ plane, for higher velocities (b), it is exclusively
located around the $F(\phi, \alpha)\approx 0$ region. The insets  show the corresponding $\rho_{v_1,v_2}(\phi,\alpha)$.} \label{fig:fig3}
\end{figure}

The effect of the population of the region around the separatrix becomes even more obvious when
considering the phase space visiting probability  velocity resolved rather than
collision resolved, $\rho_{v_1,v_2}(F,\xi)=\frac{1}{n_2}\int_{v_1}^{v_2} \sum_n \rho(F,\xi,v,n)dv$.
All collisions between $n_1=1$ and $n_2=10^7$ with
$v_1<v<v_2$ are projected onto the $F\times\xi$ space. For low velocities,
$v_1=0,\,v_2=1$ (see Fig.~\ref{fig:fig3}a), the complete $F\times\xi$ space and thus the complete $\phi\times\alpha$ plane (see the inset where the corresponding $\rho_{v_1,v_2}(\phi,\alpha)$ is shown)
is populated. For higher velocities, $v_1=30,\,v_2=40$ (see
Fig.~\ref{fig:fig3}b and the inset), a narrow, sharply bounded region around $F(\phi,\alpha)=0$ is now exclusively populated, i.e., outside this region no collisional events occur. This kind of population of
phase space is generic and does not depend on the initial
ensemble, as long as the initial velocity $v_0$ of the ensemble is sufficiently small. Unlike in dissipative systems the above squeezing of the ensemble is not caused by the existence of attractors but is due to a mechanism which will be explored in the following.

For low $v$ we encounter a  mixed
phase space, where an infinite hierarchy of  regular islands is embedded into a
large chaotic sea. With increasing $v$, large regular regions emanate (we confirmed the existence of these regions among others by calculating the corresponding Lyapunov exponents and determined their extension) from the
position of the elliptic fixed points of the static system. Their extension  in
the $\phi \times \alpha$ subspace grows rapidly with increasing  $v$ until
$v\approx 10$. From there on the  regular regions capture a substantial
amount of the available phase space volume, except a small region
around $F(\phi,\alpha)\approx0$. If we now increase $v$ further, the growth
of these regular regions is reduced significantly, leaving the channel
$F(\phi,\alpha)\approx0$  open for diffusive processes. The precondition to use sufficiently low
$v_0$ in order to observe FA becomes obvious now: It ensures that particles start inside the chaotic
sea which is connected to the channels $F(\phi,\alpha)\approx0$  that eventually
allow for the acceleration process. 

The evolution of the phase space density can then be understood as follows. The pronounced
localization of the ensemble with respect to the $F\times\xi$ and correspondingly to the $\phi\times\alpha$ space in the $v$ resolved
picture (Fig.~\ref{fig:fig3}) is smeared out when considering the collision resolved
density (Fig.~\ref{fig:fig2}). This happens because the increase of the ensemble average $\langle v\rangle$
with increasing $n$ comes along with a certain distribution
$\rho(v)$ (a Maxwell-Boltzmann like distribution \cite{Petri:2008}), i.e., the
$n$ resolved distribution is a weighted superposition of the $v$ resolved
$\rho_{v_1,v_2}(F,\xi)$, where the weights are given by $\rho(v)$.  
\begin{figure}[htp]
\centerline{\includegraphics[width=\columnwidth]{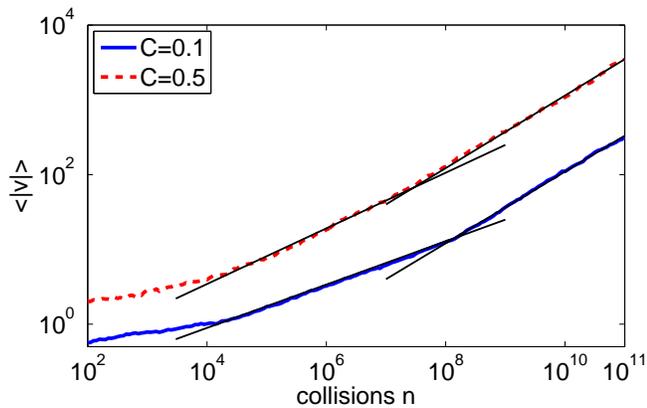}}
\caption{(Color online) Ensemble averaged velocity $\langle |\bm{v}| \rangle(n)\sim n^{\beta(C)}$ as a function
of  $n$ (collisions) for two different amplitudes $C$. Note the
different slopes $\beta$ between $n_1=3\cdot10^4-10^8$ and $n_2=10^8-10^{11}$: in the interval $n_1$  $\beta (0.1)=0.28, \, \beta (0.5)=0.40$ (subdiffusion), in the interval $n_2$ $\beta (0.1)=\beta (0.5)=0.5$ (normal diffusion).}
\label{fig:fig4}
\end{figure}

The evolution of the composition of phase space with increasing $v$ has amazing consequences on the  mean   velocity $\langle
|\bm{v}| \rangle$ of the ensemble, in particular on its long time behavior. The rich phase
space structure, associated
with many spacious regular islands embedded into the low $v$ chaotic sea, causes stickiness of the trajectories. As a result the dynamics is intermittent, exhibiting both laminar phases of oscillations as well as chaotic phases. The chaotic phases of the trajectories lead to 
stochastic fluctuations of $v$ and thus to a net
increase of the mean velocity $\langle |\bm{v}| \rangle$ of the ensemble. This intermittent interplay of laminar and turbulent phases can be
effectively described through continuous time random walks \cite{Montroll:1965} predicting
the appearance of subdiffusion, which is exactly observed here. With increasing $\langle |\bm{v}| \rangle$, the ensemble is more and more squeezed into the acceleration
channels around $F=0$.  A detailed analysis  based on the periodic orbits
shows that there are much fewer and at the same time smaller regular islands in
the acceleration channels for higher compared to low velocities, see also Fig.~\ref{fig:fig1}. This leads to a significantly reduced stickiness, i.e., fewer and shorter laminar phases once the ensemble is predominantly located
in the narrow acceleration channels.  Consequently, the diffusion in $v$ space becomes then
normal. The crossover at around $n_c\approx 10^8$ from subdiffusion to normal
diffusion can be  seen in Fig. \ref{fig:fig4} for two different values of the
amplitude ($C=0.1$ and $C=0.5$). Such a crossover  occurs in other driving modes of the ellipse as well, e.g. in the quadrupole mode, where $a(t)=a_0+C\sin \omega t$ and $b(t)= b_0-C\sin \omega t$ \cite{Petri:2008}. In Fig.~\ref{fig:fig4}, the average velocity grows according to a
power law $\langle |\bm{v}| \rangle(n) \sim n^{\beta(C)}$ between
$n\approx5\cdot10^4$ and $n\approx 10^8$. In this region, the diffusion exponent $\beta(C)$ is
amplitude dependent ($\beta(0.1)=0.28, \, \beta(0.5)=0.40$) and always smaller than $0.5$ (subdiffusion). The dependence of  $\beta(C)$ in this
 regime can also be traced back to the phase space properties, e.g.
there are more islands causing stickiness  for smaller  than for
larger amplitudes \cite{Petri:2008}. For $n>10^8$ $\langle |\bm{v}| \rangle(n) \sim n^{1/2}$, i.e.,
the exponent obeys $\beta=1/2$ independent of the amplitude. Note however that an
ensemble of particles has to be propagated for $n\gg n_c\approx10^8$ (until $n=10^{11}$ in
our case), which is extremely demanding and requires large scale
computations, to detect this crossover. We checked the convergence of the ensemble with 160 particles shown in Fig.~\ref{fig:fig4} with a reference ensemble consisting of $10^4$ particles and shorter iteration times and found perfect agreement. 

Our analysis of the 4D phase space of the driven elliptical billiard shows how its composition dynamically changes with increasing velocity $v$. For high $v$, thin acceleration channels of diffusive motion evolve from a spacious chaotic sea located at low $v$. This induces a crossover from sub to normal diffusion  in momentum space. The acceleration channels are present in particular in the high $v$ quasistatic limit. We expect this to be a generic feature of many driven dynamical systems which show Fermi acceleration. With increasing $v$, the ensemble will be focused on phase space regions  where the corresponding  static system possesses the highest degree of instability, or even chaoticity in the case of non-integrable billiards. In this sense driven billiards, as shown in the case of elliptical
geometry,  may operate not only as an accelerator but also as a phase
space collimator.

Financial support by the Deutsche Forschungsgemeinschaft is gratefully acknowledged. F.L. acknowledges support from
the Landesgraduiertenf\"orderung Baden-W\"urttemberg in the framework of the Heidelberg Graduate School of Fundamental Physics (HGSFP). F.K.D. likes to thank the HGSFP for financial support.

\end{document}